\documentclass[traditabstract]{aa} 

\usepackage{graphicx}
\usepackage{caption}
\usepackage{subcaption}
\usepackage{epstopdf}
\usepackage{amsmath}
\usepackage[english]{babel}
\usepackage{amssymb}
\usepackage{wrapfig}
\usepackage{pdflscape}
\usepackage{lscape}
\usepackage{longtable}
\usepackage{appendix}
\usepackage{textcomp}
\usepackage{multirow}
\bibpunct{(}{)}{;}{a}{}{,}
\usepackage{array}
\usepackage{ragged2e}
\usepackage{adjustbox}
\usepackage{txfonts}

\begin{document} 

   \title{Spectacular 240~kpc double-sided relativistic jets in a spiral-hosted narrow-line Seyfert 1 galaxy}
\titlerunning{Spectacular jet in an NLS1}

   \author{A. Vietri\inst{1}\thanks{\email{amelia.vietri@phd.unipd.it}}
            \and
            E. J\"{a}rvel\"{a}\inst{2}
           \and
          M. Berton\inst{3}
          \and 
          S. Ciroi\inst{1}
          \and
          E. Congiu\inst{4}
          \and
          S. Chen\inst{5}
          \and
          F. Di Mille\inst{6}
      }

\authorrunning{A. Vietri et al.}

   \institute{Dipartimento di Fisica e Astronomia “G. Galilei”, Universit{\`a} di Padova, Vicolo dell’Osservatorio 3, 35122 Padova, Italy
         \and
           European Space Agency, European Space Astronomy Centre, C/ Bajo el Castillo s/n, 28692 Villanueva de la Cañada, Madrid, Spain;
        \and
        European Southern Observatory (ESO), Alonso de C\'ordova 3107, Casilla 19, Santiago 19001, Chile;
        \and
       Departamento de Astronomía, Universidad de Chile, Camino del Observatorio 1515, Las Condes, Santiago, Chile;
        \and
        Department of Physics, Technion 32000, Haifa 32000, Israel;
        \and
        Las Campanas Observatory - Carnegie Institution for Science, Colina el Pino, Casilla 601, La Serena, Chile.}
   \date{Received ; accepted }


  \abstract
   {Narrow-line Seyfert 1 (NLS1) galaxies are a peculiar subclass of active galactic nuclei (AGN). They have demonstrated that the presence of relativistic jets in an AGN is not strictly related to its radio-loudness, the black hole mass, or the host galaxy type. Here we present a remarkable example of a radio-quiet NLS1, 6dFGS gJ035432.8-134008 (J0354-1340). In our Karl G. Jansky Very Large Array observations at 5.5~GHz the source shows a bright core with a flat spectral index, and extended emission corresponding to very elongated jets. These are the largest double-sided radio jets found to date in an NLS1, with a deprojected linear size of almost 250~kpc. We also analysed near-infrared and optical images obtained by the Magellan Baade and the European Southern Observatory New Technology Telescope. By means of photometric decomposition and colour maps, we determined that J0354-1340 is hosted by a spiral/disk galaxy. Fully evolved relativistic jets have traditionally been associated with high-mass elliptical galaxies hosting the most massive black holes, but our results confirm that also less massive black holes in spiral galaxies can launch and sustain powerful jets, implying that the launching of the jets is governed by factors other than previously believed.}
   

   \maketitle


\section{Introduction}
\label{sec:intro}
Described for the first time by \citet{1985osterbrock1}, narrow-line Seyfert 1 (NLS1) galaxies are a peculiar subclass of active galactic nuclei (AGN). Unlike broad-line Seyfert 1 galaxies (BLS1s), NLS1s are characterised by narrow permitted lines \citep{2011pogge1}. By definition, they have a small full-width at half maximum, \textit{FWHM}(H$\beta)$ $<2000$ km s$^{-1}$, \citep{1989goodrich1} and a flux ratio of [O III]$\lambda$5007/H$\beta<3$ \citep{1985osterbrock1}.
Unlike in Type 2 AGN, the narrowness of the permitted lines is not related to obscuration. The low [O III]$\lambda$5007/H$\beta<3$ ratio and the presence of strong Fe~II multiplets in many NLS1s indicate an unobscured view of the broad-line region (BLR).

If the broadening of the permitted lines is due to Keplerian motion around a black hole (BH), their spectral features could be explained as a consequence of a low rotational velocity around a low-/intermediate-mass BH ($M_{\rm BH}<10^{8}$ $M_{\odot}$, \citealp{2011peterson1}). Considering also that the bolometric luminosity, $L_{\rm bol}$, of these AGN is comparable to that of BLS1s, it is inferred that a fraction of NLS1s accrete close to, or even above, the Eddington limit \citep{1992boroson1}. Some authors have, instead, proposed that the narrowness of the permitted lines in NLS1s is due to an inclination effect, caused by the lack of Doppler broadening, due to the pole-on view of a disk-like BLR \citep{2008decarli1}.

Whereas reverberation mapping has confirmed the low BH masses in some NLS1s \citep{2011peterson1}, it is arduous and time-consuming to obtain reverberation data for large samples. Instead, their host galaxy morphologies can be used as a first-order estimate of the BH mass to investigate whether they actually are low. The host galaxy of an AGN interacts and co-evolves with the nuclear region, and thus the host galaxy morphology is linked to the properties of the central engine \citep{2000ferrarese1, Morganti17}. There are relations between the BH mass and properties of the bulge of its host galaxy \citep{Magorrian98}, and, generally, more massive BHs are found to be hosted in elliptical galaxies, while spiral and disk galaxies tend to harbour less massive BHs \citep{Salucci00}. As predicted by the low-mass scenario, it has been found that NLS1s, also the jetted ones, are preferably hosted by disk-like galaxies, often with pseudo-bulges and bars \citep{2001krongold1, 2003crenshaw1, 2006deo1, 2008anton1, 2011orbandexivry1, 2012mathur1, 2016kotilainen1, 2017olguiniglesias1, 2017dammando1, 2018dammando1, Jarvela18b, 2019berton1, Hamilton21}. Due to this ensemble of properties NLS1s are thought to be unevolved AGN \citep{2001mathur1}.

Despite multiple observations suggesting NLS1s having low-mass BHs, several NLS1s have been detected at $\gamma$-rays \citep{2009abdo2, 2009abdo1, 2018paliya1, Foschini21}, confirming the presence of powerful relativistic jets in them. This goes against the fact that, until recently, it was believed that powerful jets could only be harboured in elliptical galaxies, hosting the most massive supermassive BHs  \citep{1995urry1, 1998franceschini1, Kotilainen05}. This result brings new evidence to support the idea that spiral galaxies with pseudo-bulges and low-mass BHs can launch and sustain powerful jets \citep{2011foschini1, 2020olguiniglesias1}. Mergers and interaction, which are more common in crowded regions, could play a role in triggering the launching of relativistic jets \citep{2015chiaberge1}. Jetted NLS1 galaxies residing in denser large-scale environments than non-jetted NLS1 galaxies support this scenario \citep{2017jarvela1}.

The presence of jets in NLS1s is not strictly related to their radio-loudness parameter \textit{R}\footnote{Radio-loudness \textit{R} is defined as the ratio between the radio 5~GHz flux density and the optical B-band flux density,  \citep{1989kellermann1}}. In only $\approx$7$\%$ of them the radio emission clearly dominates over the optical emission \citep{2006komossa1}, but even these sources do not necessarily harbour jets \citep{2015caccianiga1}. On the other hand, some NLS1s with no prior radio detections at any frequency, have recently been detected at 37~GHz, confirming the presence of jets in them \citep{2018lahteenmaki1, 2020berton1, 2021jarvela2}.

The radio morphologies of NLS1s are diverse, and a fraction of them have been found to host large kiloparsec-scale jets \citep[e.g.,][]{2018berton1,Jarvela22}. Possibly, the most remarkable example is the source we study in this work, that shows very extended two sided relativistic jets, and was discovered by \citet{2020chen1}.
In this letter, we study this source in more detail by investigating its radio emission, and the properties of its host galaxy. This work is organised as follows. In Sect.~\ref{sec:j0354} we describe the source, in Sect.~\ref{sec:radio} we present the radio analysis, Sect.~\ref{sec:host} is dedicated to the near-infrared (NIR) and optical analysis of the host galaxy, while in Sect.~\ref{sec:disc} we discuss the results and give conclusions.

Throughout this study, we adopt a standard $\Lambda$CDM cosmology, with a Hubble constant $H_{0} = 67.8$ km s$^{-1}$ Mpc$^{-1}$, considering a flat Universe with the matter density parameter $\Sigma_{M} = 0.308$ and the vacuum density parameter $\Sigma_{vac} = 0.692$ \citep{Planck16}. For spectral indices, we adopt the convention of $S_{\nu}$ $\propto$ $\nu^{\alpha}$ at frequency $\nu$.


\section{6dFGS gJ035432.8-134008}
\label{sec:j0354}
The target of this study is the NLS1 6dFGS gJ035432.8-134008 (hereafter J0354-1340, R.A. 03:54:32.85, Dec. -13:40:07.24) at $z = 0.076$. This source is classified as an NLS1 by \citet{2018chen1}, using a Six-degree Field Galaxy Survey  spectrum. On the basis of the H$\beta$ emission line dispersion and the continuum luminosity at 5100~$\AA$, its BH mass is estimated to be $M_{\mathrm{BH}} \approx 9.8 \times 10^{6} M_{\odot}$. The radio morphology of J0354-1340 was studied by \citet{2020chen1} who classified this source as a Fanaroff-Riley II (FR~II) radio galaxy \citep{1974fanaroff1}. It has a compact central core, probably the jet-base, with a flat in-band spectral index. This NLS1 also shows considerable extended emission at both sides of the core at kiloparsec-scales. \citet{2020chen1} report the projected extent of the southern radio emission to be 93.5~kpc, and of the northern to be 83.9~kpc, and measure the total integrated luminosity of $\nu L_{\nu, \mathrm{int}} \approx 3.8 \cdot 10^{39}$ erg  s$^{-1} $ at 5.5~GHz. It should be noted that despite its unusual radio properties, the radio emission of J0354-1340 is not considerably brighter than its optical emission \citep{2020chen1}.


\section{Radio morphology}
\label{sec:radio}
\begin{figure}
    \centering
    \includegraphics[width=\hsize]{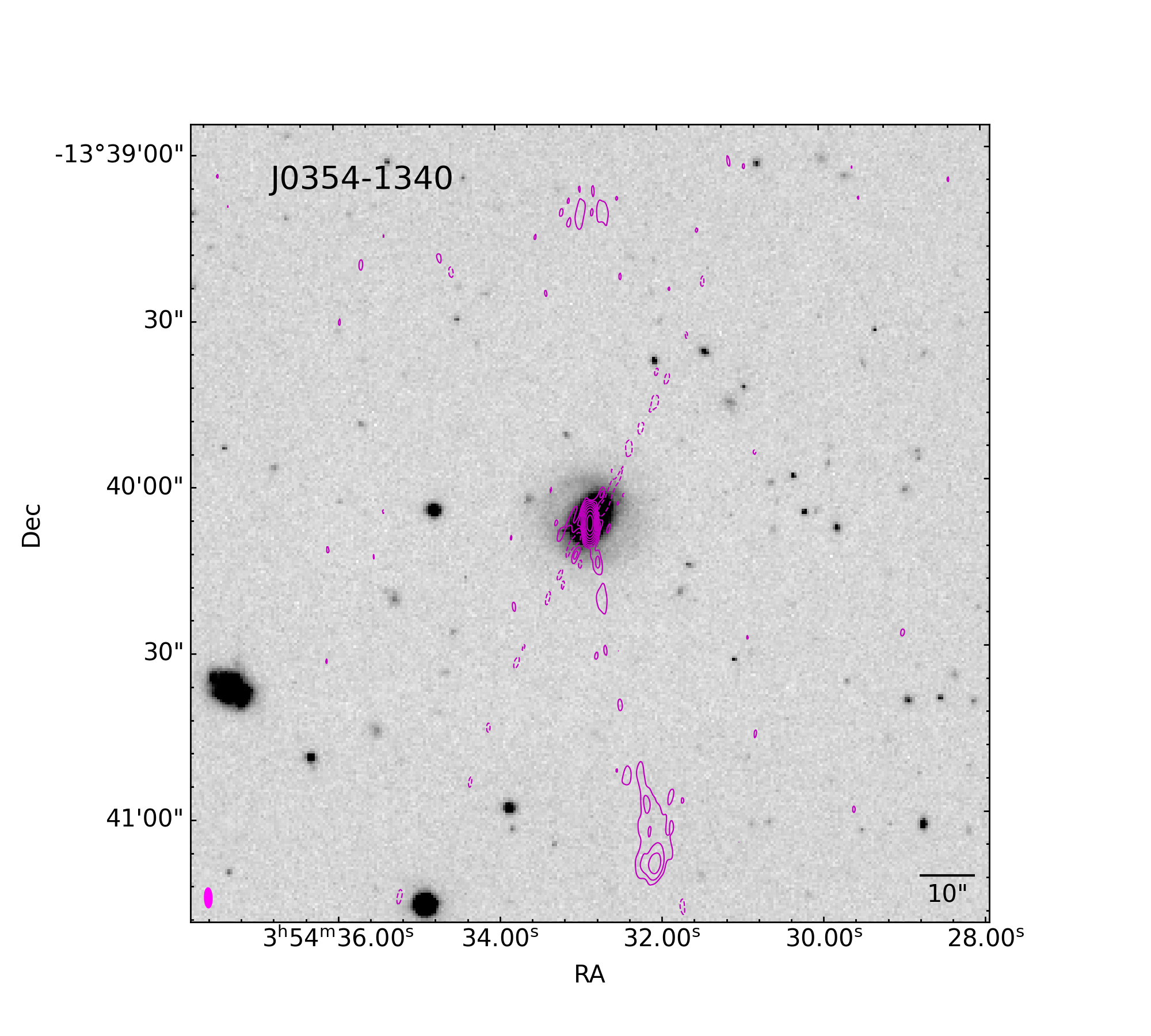}
    \caption{Radio map of J0354-1340 overlaid with the $J$-band host galaxy image. The rms of the radio map is 8~$\mu$Jy beam$^{-1}$. The relative contour levels are at rms $\times$ -3, 3 $\times$ 2$^n$ n $\in$ [0, 7]. The beam size is 5.25~kpc $\times$ 1.93~kpc.}
    \label{fig:radiomap}
\end{figure}

\begin{figure}
    \centering
    \includegraphics[width=\hsize]{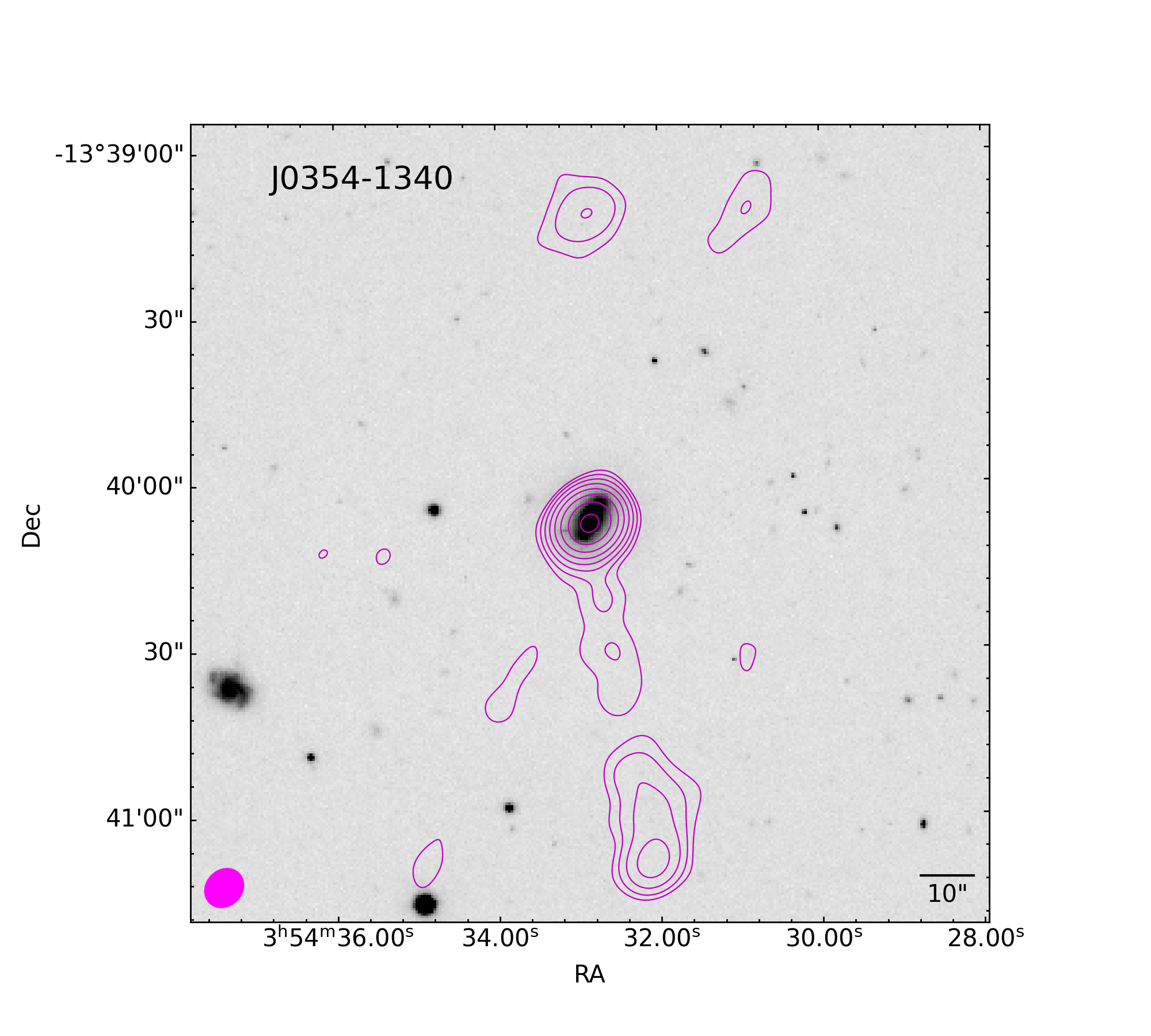}
    \caption{Tapered map of J0354-1340 overlaid with the $J$-band host galaxy image. The relative contour levels are at rms $\times$ -3, 3 $\times$ 2$^n$ n $\in$ [0, 7]. The rms is 11~$\mu$Jy beam$^{-1}$. The beam size is 11.10~kpc $\times$ 9.50~kpc.}
    \label{fig:tapmap}
\end{figure}

\begin{figure}
    \centering
    \includegraphics[width=\hsize]{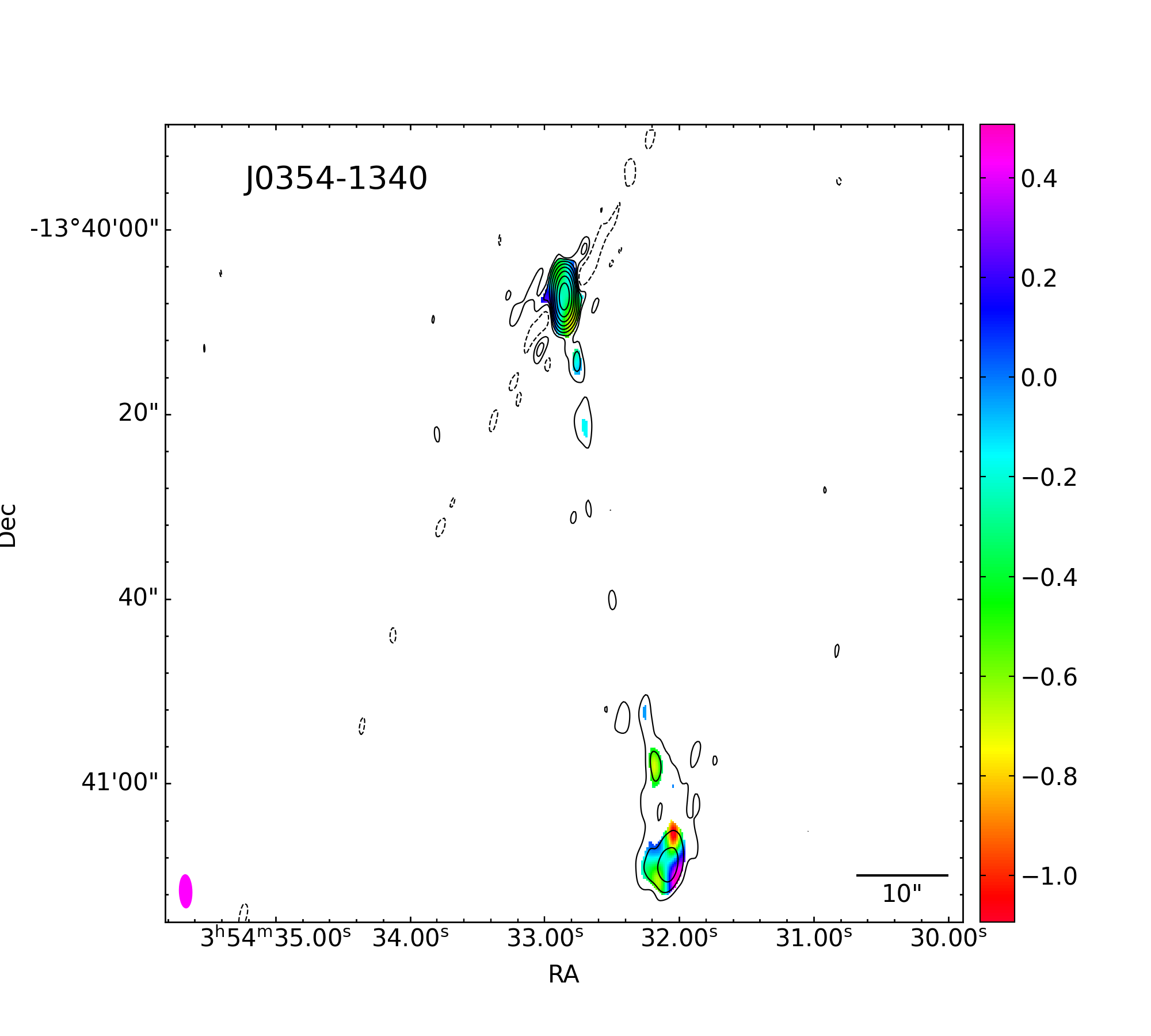}
    \caption{Spectral index map of J0354-1340, showing only the core and the southern emission. The signal-to-noise of the northern emission is too low  to calculate its spectral index. The beam size is the same as in Fig.~\ref{fig:radiomap}.}
    \label{fig:alphamap}
\end{figure}

\begin{figure}
    \centering
    \includegraphics[width=\hsize]{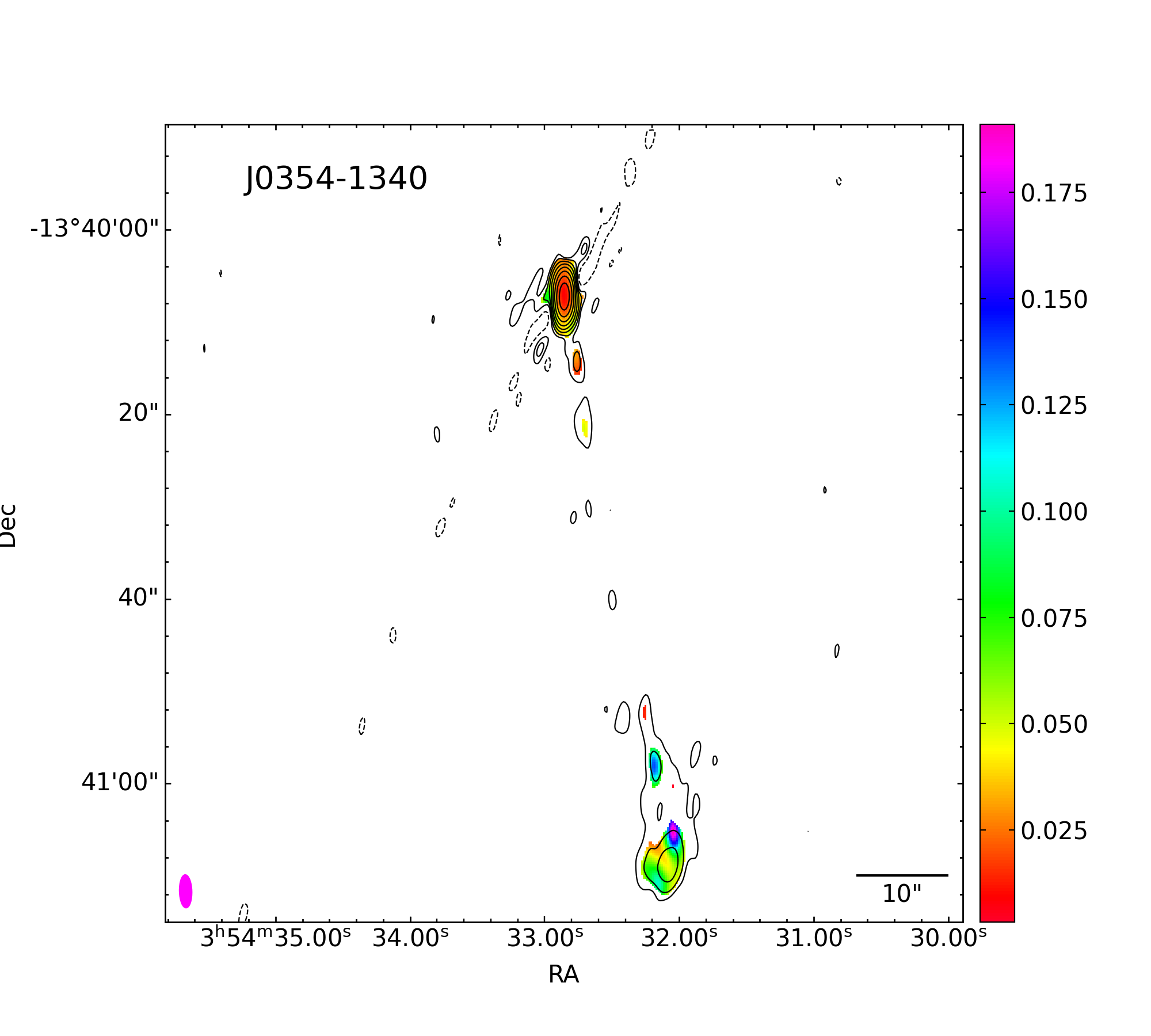}
    \caption{The error spectral index map of J0354-1340.}
    \label{fig:alphaerrmap}
\end{figure}

In this work, we use the same Karl G. Jansky Very Large Array (JVLA) data as in \citet{2020chen1}, but we perform a more detailed analysis. The radio data were obtained in February 2019 with the JVLA in C-configuration, centred at 5.5~GHz with a bandwidth of 2~GHz, divided into 16 spectral windows. The exposure time was 30~minutes and the angular resolution 3.5", yielding a theoretical image sensitivity of $\sim 7 \mu$Jy beam$^{-1}$. The data were calibrated using the VLA pipeline version 5.4.0. For further data reduction and analyses we used Common Astronomy Software Applications (CASA) version 6.1.2-7. The data of J0354-1340 were split from the measurement set and averaged in time (10~sec) and frequency (64 channels). To obtain the radio map, the tapered map, and the spectral index map, we used the multi-term multi-frequency synthesis, \texttt{mtmfs}, method implemented in CASA, and followed the procedure by \citet{Wiegert15} and \citet{Jarvela22}. The \texttt{mtmfs} algorithm models the wide-band sky-brightness distribution as a linear combination of spatial and spectral basis functions, and performs image-reconstruction \citep{Rau_Cornwell11}.
The radio map, overlaid with the $J$-band host galaxy image (Fig.~\ref{fig:radiomap}), shows extended emission, corresponding to very elongated jets, located on the south-west, and the northern sides of the nucleus. We hypothesise the northern emission is real, albeit being detected only at 3~$\sigma$, because the archival National Radio Astronomy Observatory VLA Sky Survey radio map shows extended emission toward north. Since J0354-1340 shows extended structures, we convolved the visibilities with a Gaussian taper with a FWHM of 15~k$\lambda$, to enhance the extended emission sensitivity. The tapered map, overlapped on the $J$-band image of the source, is shown in Fig.~\ref{fig:tapmap}. However, the tapered map does not reveal new regions of radio emission. The spectral index map (Fig.~\ref{fig:alphamap}) shows a flat core, as typically seen in blazars. The spectral index of the lobe is approximately -0.5, which is slightly higher than the characteristic spectral index of -0.7 of optically thin synchrotron emission. CASA also provides the empirical error associated with the spectral indices, as shown in Fig.~\ref{fig:alphaerrmap}.

We measured the peak flux density and its error, produced by CASA, by fitting a 2D Gaussian to the core.
We also measured the core and the extended emission flux densities within the 3~$\sigma$ contour. The errors related to the flux densities were estimated as the rms ($\sim$8 $\mu$Jy) multiplied by the square root of the area covered by the outermost contour expressed in units of beam \citep{2018berton1}. All the measurements with their respective errors are reported in Tab.~\ref{tab:tab1}.

\begin{table}[ht!]
\caption[]{Radio properties of the source}
\centering
\begin{tabular}{l l}
\hline\hline
Region  & Flux density \\ \hline
Core, peak & 5.25 $\pm$  0.02  mJy beam$^{-1}$ \\
Core, total & 4.88 $\pm$ 0.02  mJy \\
North, total & 0.09 $\pm$ 0.01 mJy \\
South, total & 0.76 $\pm$ 0.03 mJy  \\
Total & 5.73 $\pm$ 0.04  mJy\\ \hline
\end{tabular}
\tablefoot{Columns: (1) Region; (2) Flux density.}
\label{tab:tab1}
\end{table}

Furthermore, we estimated the deprojected linear size, the orientation, and the age of the jet. We estimated the inclination using two different speeds for the jet: somewhat unrealistic v$_{jet}$ = 0.99~$c$, which gives an upper limit for the inclination, and a more realistic v$_{jet}$ = 0.5~$c$ \citep[e.g.,][]{Giroletti09}. Since we were able to detect emission also from the receding jet, we can use the flux density ratio of the approaching ($f_+$) and the receding jet ($f_-$) to estimate the inclination, $\theta$, that is the angle with respect to the line of sight, using \citep{2012beckmann1}:

\begin{equation}
    \frac{f_{+}}{f_{-}}= \left( \frac{1 + \beta \cdot cos(\theta)}{1 - \beta \cdot cos(\theta)}\right)^{(2 - \alpha)} \,  ,
\label{eq:estimate1}
\end{equation}

where $\alpha$ is the spectral index of the jet, set to $-1$. We chose the standard value of $\alpha$ since no reliable estimate was derived from the spectral index map. The flux densities were $f_{+}=0.76$ mJy and $f_{-}= 0.09 $ mJy, for the approaching and the receding jet, respectively. Assuming $\beta = 0.99$, Eq.~\ref{eq:estimate1} gives an estimate of the jet inclination of $\theta_1 = 70^{\circ}$. This value represents an upper limit for the inclination of the jet. Using $\beta = 0.5$ instead, we obtained a more reasonable estimate of the jet inclination of $\theta_2 = 47^{\circ}$.

We estimated the deprojected sizes of the jets using the more reasonable assumption of $\beta = 0.5$, and obtained $D^{+}_{\mathrm{jet}}$ = 127~kpc and $D^{-}_{\mathrm{jet}}$ =114~kpc, giving a total extent of 241~kpc. Interestingly, they do not seem to be exactly aligned, but at an angle of 169$^{\circ}$.

Finally, we estimated the age of the approaching jet as the ratio between its deprojected liner size and jet velocity. For $\beta =0.99c$, the age is $t \sim$ 0.3~Ma, while for $\beta = 0.5$, we have $t \sim$ 0.8~Ma.



\section{Host galaxy}
\label{sec:host}

\subsection{Photometric decomposition}
\label{sec:modelling} 

The NIR observations of J0354-1340 were carried out in October 2019, with the Magellan Baade 6.5m telescope. The $J$ and $Ks$-band  observations were performed using the wide-area NIR camera FourStar. During the observing night, the sky was clear and the seeing was approximately $0.5"$ in both bands. The total exposure time was $\approx$ $1200$~sec and $\approx$ $760$~sec, for the $J$ and $Ks$-band images, respectively. Standard reduction, photometric and astrometric calibrations, and sky subtraction were applied to the images with the Image Reduction and Analysis Facility (IRAF) Software. We performed a photometric decomposition of the $Ks$-band image, using the 2D fitting algorithm GALFIT version 3.0.5 \citep{Peng10}. More details about the photometric decomposition and host modelling are given in the App.~\ref{app:imaging-main}. Since the AGN emission affects the host model, it needs to be modelled  simultaneously with the galaxy. Because the AGN is a point-like source, this can be done by fitting it with a modelled point-spread function (PSF). We tried different approaches to create the PSF, and after several attempts the best model was obtained by fitting a nearby star with five Sérsic functions. 

We modelled the host galaxy varying at each step the initial fitting parameters to make sure that the resulting model parameters were stable, and we also varied the number of components until the best, stable fit was achieved. The variation of the sky background is the largest source of error in the fit. We estimated the $\sigma_{sky}$ by measuring the standard deviation of the sky in several empty regions in the image and taking the average of those. There are three sources of error for the magnitude: the error related to the zero point calculation, the error reported for the star magnitude values in the 2MASS Catalogue\footnote{https://irsa.ipac.caltech.edu/Missions/2mass.html}, and the error we obtained from GALFIT by modelling the source with the $\pm \sigma_{sky}$ values. To obtain the total error, we summed in quadrature the different sources of error after converting to linear units. To estimate the errors of the other parameters, except the magnitude, we fitted the source with the $\pm \sigma_{sky}$, and used the difference between the resulting values and the best-fit values as the error. The best fit was achieved with four components: a PSF for the AGN and three Sérsic functions for the bulge (S1), for the disk (S2), and for the bar (S3). The goodness of fit was determined based on the $\chi^{2}_{\nu}$ value, 1.27, and that the output parameters remained physically reasonable and stable during the fit. The S1 component represents the bulge. However, likely due to the remarkable strength of the bar component and the bulge being only marginally resolved, our measurements of the Sérsic index of the bulge have a relatively high error (see Table~\ref{tab:galfit}). Therefore, we will not discuss its properties in the following, because the uncertainty is too high. New higher-resolution observations are needed to accurately measure its physical properties. S2 with $r_{e} =$ 8.63~kpc, $n = 0.56$, and an axial ratio = 0.83 resembles a disk. The S3 is a bar-like component with $r_{e} =$ 4.40~kpc, $n = 0.37$, and an axial ratio = 0.33.
 
\renewcommand{\arraystretch}{1.5}
\begin{table*}[ht!]
\caption[]{Best fit parameters of J0354-1340 with $\chi^{2}_{\nu}$=1.27$_{-0.01}^{+0.09}$.}
\centering
\begin{tabular}{l l l l l l l}
\hline\hline
Function   &  mag                            & $r_{e}$                         & $n$                            & axial                          & PA                            & notes   \\
             &                                 & (kpc)                           &                                & ratio                          & (\textdegree)                 &  \\ \hline
PSF          & 14.29 $\substack{+0.39\\-0.42}$ &                                 &                                &                                &                               & nucleus\\
S1 & 14.06 $\substack{+0.40\\-0.55}$ & 1.15 $\substack{+0.12\\-0.75}$ & 4.24 $\substack{+0.74\\-2.03}$ & 0.91 $\substack{+0.00\\-0.00}$ & -45.81 $\substack{+0.56\\-5.41}$ & bulge \\
S2 & 13.72 $\substack{+0.42\\-0.40}$ & 8.63 $\substack{+0.79\\-2.89}$  & 0.56 $\substack{+0.08\\-2.01}$ & 0.83 $\substack{+0.04\\-0.07}$ & -38.94 $\substack{+1.69\\-0.0.}$ & disk \\
S3 &  14.51$\substack{+0.39\\-0.60}$ & 4.40 $\substack{+0.08\\-0.33}$  & 0.37 $\substack{+0.01\\-0.12}$ & 0.33 $\substack{+0.01\\-0.06}$ & -28.24 $\substack{+0.02\\-0.89}$ & bar \\ \hline
\end{tabular}
\tablefoot{Columns: (1) Function used in the model; (2) Magnitude of the component in $Ks$-band; (3) Effective radius; (4) S\'{e}rsic index; (5) Axial ratio; (6) Position angle; (7) Physical interpretation.}
\label{tab:galfit}
\end{table*} 

\begin{figure}
    \centering
    \includegraphics[width=9cm]{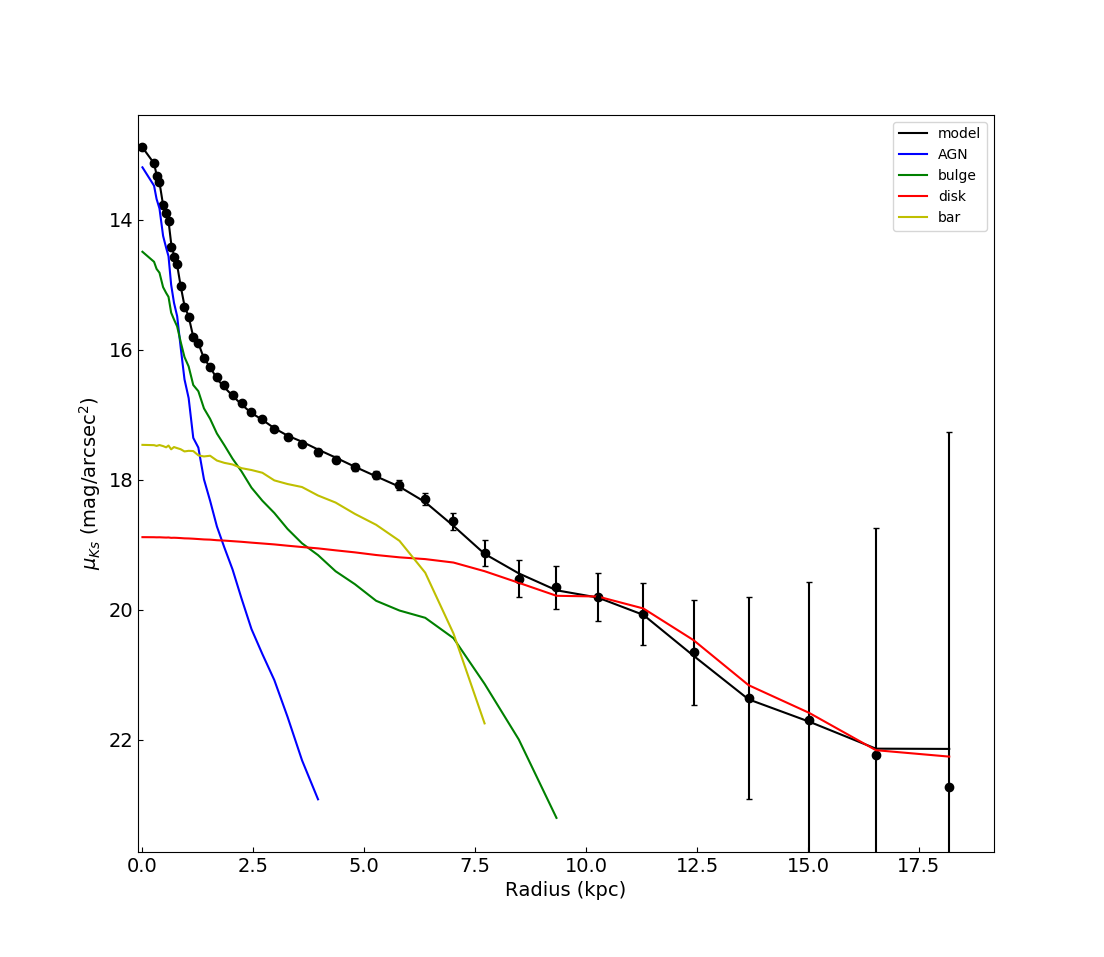}
    \caption{Radial surface brightness profile of the galaxy. The image shows the model in black, the AGN component in blue, the bugle component in dark green, the disk component in red, the bar component in light green.}
    \label{fig:profile}
\end{figure}

\begin{figure*}
    \centering
    \includegraphics[width=18cm]{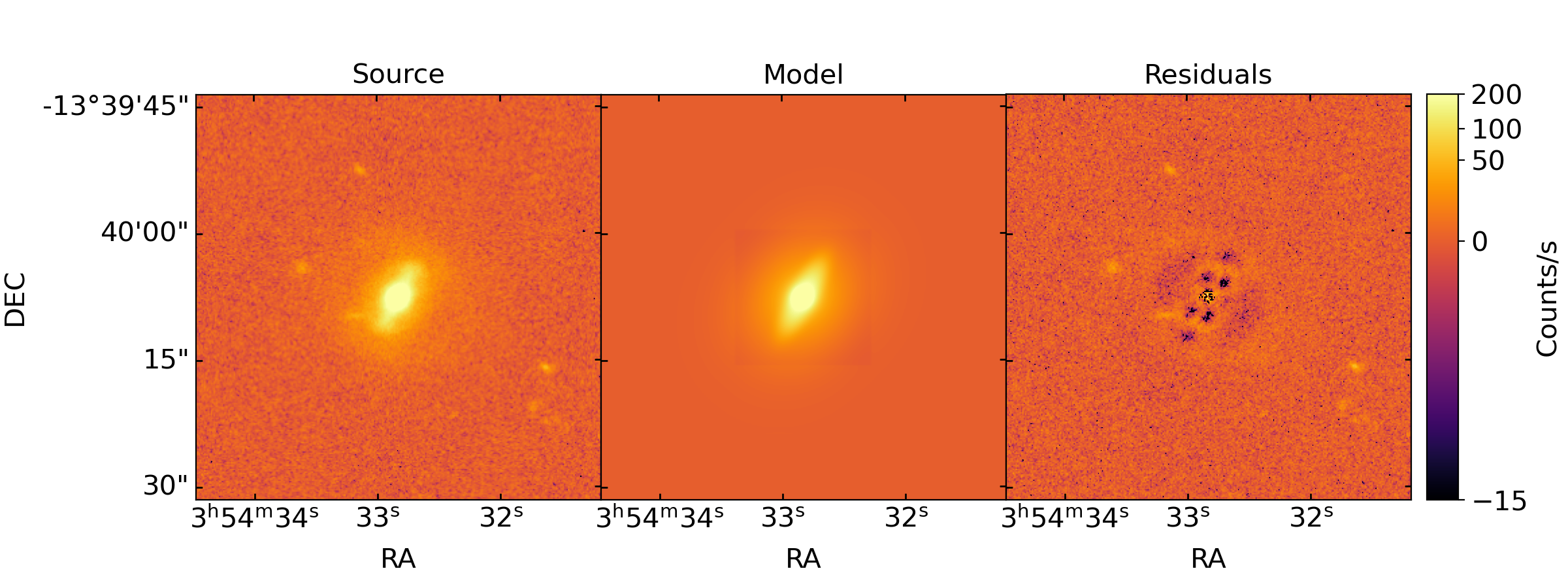}
    \caption{\textbf{Left panel}: observed image in $Ks$-band. The size of the image, the one of the fitting region, is 47.7 arcsec, corresponding to 70.9 kpc. \textbf{Central panel}: model image of the galaxy. \textbf{Right panel}: residual image, smoothed over 2~px.}
    \label{fig:plot3}
\end{figure*}

We found out that this NLS1 is hosted in a barred spiral/disk galaxy while no further details can be given about the characterisation of the bulge, since it is not properly resolved. In the observed image, shown in the left panel of Fig.~\ref{fig:plot3}, it is possible to see the clear barred spiral shape, with a peculiar bar enhanced at the poles and a faint ring surrounding the whole galaxy. This galaxy shows a very complex structure but since we are mainly interested in the main components of the galaxy we did not try to model the more complex features, such as the faint ring. The image shows also some small unmodelled regions outside the galaxy which also affect the $\chi^{2}_{\nu}$ value, that is calculated for the whole fitting region. The model is shown in the central panel of Fig.~\ref{fig:plot3}. The residuals, in the right panel of Fig.~\ref{fig:plot3}, show a ring-like structure around the centre, especially enhanced at the "poles" of the bar, while in the centre there are both over- and under-subtracted regions. 
The radial surface brightness profile of the galaxy and the galfit components, obtained with $ELLIPSE$ in IRAF, is shown in Figure.~\ref{fig:profile}. The model matches the galaxy profile well, however, the disk component is unusually faint.


\subsection{Colour images}
\label{sec:colour}
The optical images of this source were obtained with the New Technology Telescope (NTT), (Proposal ID: NTT/106.21HS, PI M. Berton) in January 2021. The $g$- and $i$-band observations were performed using the ESO Faint Object Spectrograph and Camera version 2. The seeing was about 1.13" for $g$ and 1.08" for $i$-band, while the exposure time is 300s for both images. We performed a standard reduction using IRAF, with bias and flat-field correction, followed by the alignment, sky subtraction, fringing removal, and combination of the images in each filter.

\begin{figure}
    \centering
    \includegraphics[width=9cm]{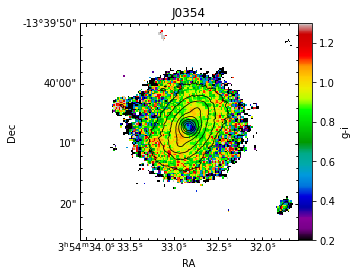}
    \caption{$g-i$ colour map of J0354-1340.}
    \label{fig:colour1}
\end{figure}

The $g-i$ colour map is shown in Fig.~\ref{fig:colour1}. The colour is quite uniform throughout the whole galaxy, except for the nucleus. The central region of the map is blue, showing $g-i$ $\approx$ 0.4, a value typically observed in green quasars \citep{Klindt19}.

\begin{figure}
    \centering
    \includegraphics[width=9cm]{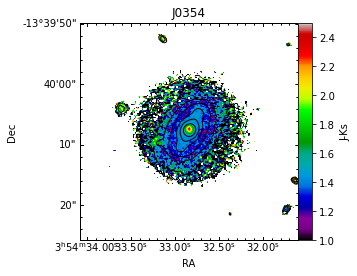}
    \caption{$J-Ks$ colour map of J0354-1340.}
    \label{fig:colour2}
\end{figure}

We also derived the $J-Ks$ colour map (Fig.~\ref{fig:colour2}). The colour is quite homogeneous and it is mainly blue; The difference between $J$ and $Ks$ magnitudes is $\sim$1.5, as would be expected for a Seyfert-like galaxy \citep{Jarrett00}. The central region colour, corresponding to the AGN, is instead red, possibly due to dust extinction or old stars. As seen in the map, the bar is really prominent, possibly due to its abundant dust content. In the NIR, also the stellar populations can contribute to the colour, especially old star that may make the colour redder.

\section{Discussion}
\label{sec:disc}

In radio, J0354-1340 shows a bright compact core and extended emission, located on the southwest and on the north side of the central region (Fig.~\ref{fig:radiomap}). These extended structures correspond to the radio lobe of the approaching and the receding jet, respectively. The presence of very extended jets in a source where the radio emission does not dominate over the optical emission is more proof that simple parameters, such as the radio-loudness, are ambiguous and should not be trusted or used \citep{2017jarvela1,2017padovani1, 2018lahteenmaki1, Berton21}. What makes this source truly peculiar among NLS1s, and other AGN with low BH masses, is the size of the extended emission. The projected linear size of 100~kpc of each jet is consistent with those of FR~II radio galaxies \citep{Hardcastle98, Kharb08}. J0354-1340 harbours indeed the largest radio jets in projected size found to date in an NLS1s, whose projected jet sizes are usually of the order of a few kiloparsecs \citep{2012doi1, 2015richards1, 2017congiu1, Rakshit18, 2018gabanyi1, Jarvela22}.

The largest estimated deprojected size of the jets among NLS1s is $\sim$1~Mpc in PMN J0948+0022 \citep{Doi19}. However, this estimate is based on an assumption of a very low inclination, as the counter-jet is not visible in radio maps. Instead, J0354-1340 has the largest deprojected linear size ever observed in an NLS1s, $\sim$ 240~kpc, with its viewing angle estimated from the jet/counter-jet flux density ratio. Naturally, also our estimate includes uncertainty due to the unknown jet speed and spectral index. The inclination derived from $v \sim c$, $\theta_1 = 70^{\circ}$, is higher than typical values for type 1 AGN \citep{Peterson97}. Therefore, the approximation $v_{jet} = 0.99c$ is most probably not realistic in this case. The more reasonable approximation of $\beta=0.5$ gives an inclination of $\theta_2 = 47^{\circ}$, which is consistent with the range of viewing angles for type 1 AGN. This value does not imply a face-on view of the nucleus and, for this reason, the BH mass should not be underestimated due to an inclination effect \citep{vietri18}. However, were the jet speed lower, it would also imply lower inclination, but probably could not have resulted in the radio morphology we see in this source. Assuming a viewing angle of $47^{\circ}$, the deprojected size of the whole extended emission is $\sim$ 241~kpc. The extent of the jets in J0354-1340 is extraordinary and unprecedented among NLS1s, who usually harbour jets that are some hundreds of parsec to a few kiloparsecs long. The jets are huge compared to the host galaxy whose half-light radius we estimated to be $\sim$8.6~kpc, and extent beyond the host galaxy, reaching the intergalactic medium (IGM). The age of the approaching jet estimated using the reasonable approximation of $\beta=0.5$ gives a lower limit of $\approx 8 \times 10^5 $ years. This lies at the low tail of the age range of $10^5$ to $10^7 $ years found for other jetted NLS1s \citep{2009czerny1}. The jets are not aligned, with an angle of 169$^{\circ}$ between them. We speculate that this is due to a combination of the motion of the galaxy through the IGM and the ram-pressure induced by the IGM \citep{plewa97, Savolainen06}.

Considering the spectral index map, the compact core, which probably represents the jet base, shows a flat in-band spectral index, as seen in blazars \citep{2015foschini1}. The spectral index of the extended emission, $\approx -0.5$, is slightly higher than the usual spectral index of $\approx -0.7$ of optically thin synchrotron emission. This may be due to interaction between the lobe and the IGM, which re-accelerates the electrons in the lobe and thus makes the spectral index flatter \citep{Best97}.

The photometric decomposition of J0354-1340 shows that it is hosted in a barred spiral galaxy. The BH mass of this source is within the typical range for NLS1 \citep{2011peterson1}, as mentioned in Sec.~\ref{sec:j0354}. This value is also consistent with the high end of the BH masses hosted in late-type spirals \citep{Salucci00}. This further confirms that fully developed relativistic jets can be harboured in spirals, as proved by several authors \citep{Keel06, Mao18, Jarvela18b, 2020olguiniglesias1}. Traditionally, the most powerful relativistic jets have been associated with the highly evolved massive elliptical galaxies with considerably higher BH masses, formed in dense large-scale environments. These galaxies undergo the phase of the formation of jets, that coincides with the formation of the classical bulge, probably induced by mergers \citep{2001kormendy1, 2003urry1}. In spiral galaxies, secular processes, rather than interaction and mergers, dominate the bulge-growth, the BH activity, and the formation of pseudo-bulges \citep{Costantin22}. Nevertheless, spiral galaxies have been found to host kiloparsec- or even megaparsec-scale relativistic jets \citep{2011hota1}. J0354-1340 lies in this scenario, harbouring one the largest radio jets found to date in an NLS1, hosted in a disk galaxy.

\begin{acknowledgements}
This paper includes data gathered with the 6.5 meter Magellan Telescopes located at Las Campanas Observatory, Chile. Based on observations made with ESO Telescopes at the La Silla Paranal Observatory under ESO programme P106.21HS. M.B. and S.C. acknowledge the financial support from the visitor and mobility program of the Finnish Centre for Astronomy with ESO (FINCA), funded by the Academy of Finland grant nr. 306531. M.B. is an ESO fellow. E.J. is a current ESA science fellow. E.C. acknowledges support from ANID project Basal AFB-170002.

\end{acknowledgements}

\bibliographystyle{aa}
\bibliography{bibliography.bib}

\begin{thebibliography}{78}
\expandafter\ifx\csname natexlab\endcsname\relax\def\natexlab#1{#1}\fi

\bibitem[{{Abdo} {et~al.}(2009{\natexlab{a}}){Abdo}, {Ackermann}, {Ajello},
  {Axelsson}, {Baldini}, {Ballet}, {Barbiellini}, {Bastieri}, {Battelino},
  {Baughman}, {Bechtol}, {Bellazzini}, {Bloom}, {Bonamente}, {Borgland},
  {Bregeon}, {Brez}, {Brigida}, {Bruel}, {Caliandro}, {Cameron}, {Caraveo},
  {Casandjian}, {Cavazzuti}, {Cecchi}, {Chekhtman}, {Cheung}, {Chiang},
  {Ciprini}, {Claus}, {Cohen-Tanugi}, {Collmar}, {Conrad}, {Costamante},
  {Dermer}, {de Angelis}, {de Palma}, {Digel}, {Silva}, {Drell}, {Dubois},
  {Dumora}, {Farnier}, {Favuzzi}, {Focke}, {Foschini}, {Frailis}, {Fuhrmann},
  {Fukazawa}, {Funk}, {Fusco}, {Gargano}, {Gehrels}, {Germani}, {Giebels},
  {Giglietto}, {Giordano}, {Giroletti}, {Glanzman}, {Grenier}, {Grondin},
  {Grove}, {Guillemot}, {Guiriec}, {Hanabata}, {Harding}, {Hartman},
  {Hayashida}, {Hays}, {Hughes}, {J{\'o}hannesson}, {Johnson}, {Johnson},
  {Johnson}, {Kamae}, {Katagiri}, {Kataoka}, {Kerr}, {Kn{\"o}dlseder}, {Kuehn},
  {Kuss}, {Lande}, {Latronico}, {Lemoine-Goumard}, {Longo}, {Loparco}, {Lott},
  {Lovellette}, {Lubrano}, {Madejski}, {Makeev}, {Max-Moerbeck}, {Mazziotta},
  {McConville}, {McEnery}, {Meurer}, {Michelson}, {Mitthumsiri}, {Mizuno},
  {Monte}, {Monzani}, {Morselli}, {Moskalenko}, {Murgia}, {Nolan}, {Norris},
  {Nuss}, {Ohsugi}, {Omodei}, {Orlando}, {Ormes}, {Paneque}, {Panetta},
  {Parent}, {Pavlidou}, {Pearson}, {Pepe}, {Pesce-Rollins}, {Piron}, {Porter},
  {Rain{\`o}}, {Rando}, {Razzano}, {Readhead}, {Reimer}, {Reimer}, {Reposeur},
  {Richards}, {Ritz}, {Rodriguez}, {Romani}, {Ryde}, {Sadrozinski}, {Sambruna},
  {Sanchez}, {Sander}, {Parkinson}, {Scargle}, {Schalk}, {Sgr{\`o}}, {Smith},
  {Spandre}, {Spinelli}, {Starck}, {Stevenson}, {Strickman}, {Suson},
  {Tagliaferri}, {Takahashi}, {Tanaka}, {Thayer}, {Thompson}, {Tibaldo},
  {Tibolla}, {Torres}, {Tosti}, {Tramacere}, {Uchiyama}, {Usher}, {Vilchez},
  {Vitale}, {Waite}, {Winer}, {Wood}, {Ylinen}, {Zensus}, {Ziegler}, {Fermi/LAT
  Collaboration}, {Ghisellini}, {Maraschi}, {Tavecchio}, \&
  {Angelakis}}]{2009abdo2}
{Abdo}, A.~A., {Ackermann}, M., {Ajello}, M., {et~al.} 2009{\natexlab{a}},
  \apj, 699, 976

\bibitem[{{Abdo} {et~al.}(2009{\natexlab{b}}){Abdo}, {Ackermann}, {Ajello},
  {Axelsson}, {Baldini}, {Ballet}, {Barbiellini}, {Bastieri}, {Baughman},
  {Bechtol}, \& et~al.}]{2009abdo1}
{Abdo}, A.~A., {Ackermann}, M., {Ajello}, M., {et~al.} 2009{\natexlab{b}},
  \apj, 707, 727

\bibitem[{{Ant{\'o}n} {et~al.}(2008){Ant{\'o}n}, {Browne}, \&
  {March{\~a}}}]{2008anton1}
{Ant{\'o}n}, S., {Browne}, I.~W.~A., \& {March{\~a}}, M.~J. 2008, \aap, 490,
  583

\bibitem[{{Beckmann} \& {Shrader}(2012)}]{2012beckmann1}
{Beckmann}, V. \& {Shrader}, C.~R. 2012, {Active Galactic Nuclei} (Wiley-VCH
  Verlag GmbH)

\bibitem[{{Berton} {et~al.}(2020){Berton}, {Bj{\"o}rklund},
  {L{\"a}hteenm{\"a}ki}, {Congiu}, {J{\"a}rvel{\"a}}, {Terreran}, \& {La
  Mura}}]{2020berton1}
{Berton}, M., {Bj{\"o}rklund}, I., {L{\"a}hteenm{\"a}ki}, A., {et~al.} 2020,
  Contributions of the Astronomical Observatory Skalnate Pleso, 50, 270

\bibitem[{{Berton} {et~al.}(2019){Berton}, {Congiu}, {Ciroi}, {Komossa},
  {Frezzato}, {Di Mille}, {Ant{\'o}n}, {Antonucci}, {Caccianiga}, {Coppi},
  {J{\"a}rvel{\"a}}, {Kotilainen}, {L{\"a}hteenm{\"a}ki}, {Mathur}, {Chen},
  {Cracco}, {La Mura}, \& {Rafanelli}}]{2019berton1}
{Berton}, M., {Congiu}, E., {Ciroi}, S., {et~al.} 2019, \aj, 157, 48

\bibitem[{{Berton} {et~al.}(2018){Berton}, {Congiu}, {J{\"a}rvel{\"a}},
  {Antonucci}, {Kharb}, {Lister}, {Tarchi}, {Caccianiga}, {Chen}, {Foschini},
  {L{\"a}hteenm{\"a}ki}, {Richards}, {Ciroi}, {Cracco}, {Frezzato}, {La Mura},
  \& {Rafanelli}}]{2018berton1}
{Berton}, M., {Congiu}, E., {J{\"a}rvel{\"a}}, E., {et~al.} 2018, \aap, 614,
  A87

\bibitem[{{Berton} \& {J{\"a}rvel{\"a}}(2021)}]{Berton21}
{Berton}, M. \& {J{\"a}rvel{\"a}}, E. 2021, Astronomische Nachrichten, 342,
  1066

\bibitem[{{Best} {et~al.}(1997){Best}, {Longair}, \& {Rottgering}}]{Best97}
{Best}, P.~N., {Longair}, M.~S., \& {Rottgering}, H.~J.~A. 1997, \mnras, 286,
  785

\bibitem[{{Boroson} \& {Green}(1992)}]{1992boroson1}
{Boroson}, T.~A. \& {Green}, R.~F. 1992, \apjs, 80, 109

\bibitem[{{Caccianiga} {et~al.}(2015){Caccianiga}, {Ant{\'o}n}, {Ballo},
  {Foschini}, {Maccacaro}, {Della Ceca}, {Severgnini}, {March{\~a}}, {Mateos},
  \& {Sani}}]{2015caccianiga1}
{Caccianiga}, A., {Ant{\'o}n}, S., {Ballo}, L., {et~al.} 2015, \mnras, 451,
  1795

\bibitem[{{Chen} {et~al.}(2018){Chen}, {Berton}, {La Mura}, {Congiu}, {Cracco},
  {Foschini}, {Fan}, {Ciroi}, {Rafanelli}, \& {Bastieri}}]{2018chen1}
{Chen}, S., {Berton}, M., {La Mura}, G., {et~al.} 2018, ArXiv: 1801.07234

\bibitem[{{Chen} {et~al.}(2020){Chen}, {J{\"a}rvel{\"a}}, {Crepaldi}, {Zhou},
  {Ciroi}, {Berton}, {Kharb}, {Foschini}, {Gu}, {La Mura}, \&
  {Vietri}}]{2020chen1}
{Chen}, S., {J{\"a}rvel{\"a}}, E., {Crepaldi}, L., {et~al.} 2020, \mnras, 498,
  1278

\bibitem[{{Chiaberge} {et~al.}(2015){Chiaberge}, {Gilli}, {Lotz}, \&
  {Norman}}]{2015chiaberge1}
{Chiaberge}, M., {Gilli}, R., {Lotz}, J.~M., \& {Norman}, C. 2015, \apj, 806,
  147

\bibitem[{{Congiu} {et~al.}(2017){Congiu}, {Berton}, {Giroletti}, {Antonucci},
  {Caccianiga}, {Kharb}, {Lister}, {Foschini}, {Ciroi}, {Cracco}, {Frezzato},
  {J{\"a}rvel{\"a}}, {La Mura}, {Richards}, \& {Rafanelli}}]{2017congiu1}
{Congiu}, E., {Berton}, M., {Giroletti}, M., {et~al.} 2017, \aap, 603, A32

\bibitem[{{Costantin} {et~al.}(2022){Costantin}, {P{\'e}rez-Gonz{\'a}lez},
  {M{\'e}ndez-Abreu}, {Huertas-Company}, {Alcalde Pampliega}, {Balcells},
  {Barro}, {Ceverino}, {Dimauro}, {Dom{\'\i}nguez S{\'a}nchez},
  {Espino-Briones}, \& {Koekemoer}}]{Costantin22}
{Costantin}, L., {P{\'e}rez-Gonz{\'a}lez}, P.~G., {M{\'e}ndez-Abreu}, J.,
  {et~al.} 2022, arXiv e-prints, arXiv:2202.02332

\bibitem[{{Crenshaw} {et~al.}(2003){Crenshaw}, {Kraemer}, \&
  {Gabel}}]{2003crenshaw1}
{Crenshaw}, D.~M., {Kraemer}, S.~B., \& {Gabel}, J.~R. 2003, \aj, 126, 1690

\bibitem[{{Czerny} {et~al.}(2009){Czerny}, {Siemiginowska}, {Janiuk},
  {Nikiel-Wroczy{\'n}ski}, \& {Stawarz}}]{2009czerny1}
{Czerny}, B., {Siemiginowska}, A., {Janiuk}, A., {Nikiel-Wroczy{\'n}ski}, B.,
  \& {Stawarz}, {\L}. 2009, \apj, 698, 840

\bibitem[{{D'Ammando} {et~al.}(2018){D'Ammando}, {Acosta-Pulido}, {Capetti},
  {Baldi}, {Orienti}, {Raiteri}, \& {Ramos Almeida}}]{2018dammando1}
{D'Ammando}, F., {Acosta-Pulido}, J.~A., {Capetti}, A., {et~al.} 2018, \mnras,
  478, L66

\bibitem[{{D'Ammando} {et~al.}(2017){D'Ammando}, {Acosta-Pulido}, {Capetti},
  {Raiteri}, {Baldi}, {Orienti}, \& {Ramos Almeida}}]{2017dammando1}
{D'Ammando}, F., {Acosta-Pulido}, J.~A., {Capetti}, A., {et~al.} 2017, \mnras,
  469, L11

\bibitem[{{Decarli} {et~al.}(2008){Decarli}, {Dotti}, {Fontana}, \&
  {Haardt}}]{2008decarli1}
{Decarli}, R., {Dotti}, M., {Fontana}, M., \& {Haardt}, F. 2008, \mnras, 386,
  L15

\bibitem[{{Deo} {et~al.}(2006){Deo}, {Crenshaw}, \& {Kraemer}}]{2006deo1}
{Deo}, R.~P., {Crenshaw}, D.~M., \& {Kraemer}, S.~B. 2006, \aj, 132, 321

\bibitem[{{Doi} {et~al.}(2012){Doi}, {Nagira}, {Kawakatu}, {Kino}, {Nagai}, \&
  {Asada}}]{2012doi1}
{Doi}, A., {Nagira}, H., {Kawakatu}, N., {et~al.} 2012, \apj, 760, 41

\bibitem[{{Doi} {et~al.}(2019){Doi}, {Nakahara}, {Nakamura}, {Kino},
  {Kawakatu}, \& {Nagai}}]{Doi19}
{Doi}, A., {Nakahara}, S., {Nakamura}, M., {et~al.} 2019, \mnras, 487, 640

\bibitem[{{Fanaroff} \& {Riley}(1974)}]{1974fanaroff1}
{Fanaroff}, B.~L. \& {Riley}, J.~M. 1974, \mnras, 167, 31P

\bibitem[{{Ferrarese} \& {Merritt}(2000)}]{2000ferrarese1}
{Ferrarese}, L. \& {Merritt}, D. 2000, \apjl, 539, L9

\bibitem[{{Foschini}(2011)}]{2011foschini1}
{Foschini}, L. 2011, in Narrow-Line Seyfert 1 Galaxies and their Place in the
  Universe, \url{https://pos.sissa.it/126/024/pdf}

\bibitem[{{Foschini} {et~al.}(2015){Foschini}, {Berton}, {Caccianiga}, {Ciroi},
  {Cracco}, {Peterson}, {Angelakis}, {Braito}, {Fuhrmann}, {Gallo}, {Grupe},
  {J{\"a}rvel{\"a}}, {Kaufmann}, {Komossa}, {Kovalev}, {L{\"a}hteenm{\"a}ki},
  {Lisakov}, {Lister}, {Mathur}, {Richards}, {Romano}, {Sievers},
  {Tagliaferri}, {Tammi}, {Tibolla}, {Tornikoski}, {Vercellone}, {La Mura},
  {Maraschi}, \& {Rafanelli}}]{2015foschini1}
{Foschini}, L., {Berton}, M., {Caccianiga}, A., {et~al.} 2015, \aap, 575, A13

\bibitem[{{Foschini} {et~al.}(2021){Foschini}, {Lister}, {Ant{\'o}n}, {Berton},
  {Ciroi}, {March{\~a}}, {Tornikoski}, {J{\"a}rvel{\"a}}, {Romano},
  {Vercellone}, \& {Dalla Bont{\`a}}}]{Foschini21}
{Foschini}, L., {Lister}, M.~L., {Ant{\'o}n}, S., {et~al.} 2021, Universe, 7,
  372

\bibitem[{{Franceschini} {et~al.}(1998){Franceschini}, {Vercellone}, \&
  {Fabian}}]{1998franceschini1}
{Franceschini}, A., {Vercellone}, S., \& {Fabian}, A.~C. 1998, \mnras, 297, 817

\bibitem[{{Gab{\'a}nyi} {et~al.}(2018){Gab{\'a}nyi}, {Frey}, {Paragi},
  {J{\"a}rvel{\"a}}, {Morokuma}, {An}, {Tanaka}, \& {Tar}}]{2018gabanyi1}
{Gab{\'a}nyi}, K.~{\'E}., {Frey}, S., {Paragi}, Z., {et~al.} 2018, \mnras, 473,
  1554

\bibitem[{{Giroletti} \& {Polatidis}(2009)}]{Giroletti09}
{Giroletti}, M. \& {Polatidis}, A. 2009, Astronomische Nachrichten, 330, 193

\bibitem[{{Goodrich}(1989)}]{1989goodrich1}
{Goodrich}, R.~W. 1989, \apj, 342, 224

\bibitem[{{Graham} \& {Driver}(2005)}]{2005graham1}
{Graham}, A.~W. \& {Driver}, S.~P. 2005, \pasa, 22, 118

\bibitem[{{Hamilton} {et~al.}(2021){Hamilton}, {Berton}, {Ant{\'o}n}, {Busoni},
  {Caccianiga}, {Ciroi}, {G{\"a}ssler}, {Georgiev}, {J{\"a}rvel{\"a}},
  {Komossa}, {Mathur}, \& {Rabien}}]{Hamilton21}
{Hamilton}, T.~S., {Berton}, M., {Ant{\'o}n}, S., {et~al.} 2021, \mnras, 504,
  5188

\bibitem[{{Hardcastle} {et~al.}(1998){Hardcastle}, {Alexander}, {Pooley}, \&
  {Riley}}]{Hardcastle98}
{Hardcastle}, M.~J., {Alexander}, P., {Pooley}, G.~G., \& {Riley}, J.~M. 1998,
  mnras, 296, 445

\bibitem[{{Hota} {et~al.}(2011){Hota}, {Sirothia}, {Ohyama}, {Konar}, {Kim},
  {Rey}, {Saikia}, {Croston}, \& {Matsushita}}]{2011hota1}
{Hota}, A., {Sirothia}, S.~K., {Ohyama}, Y., {et~al.} 2011, \mnras, 417, L36

\bibitem[{{Jarrett}(2000)}]{Jarrett00}
{Jarrett}, T.~H. 2000, ASTRONOMICAL SOCIETY OF THE PACIFIC, 112, 1008

\bibitem[{{J{\"a}rvel{\"a}} {et~al.}(2021){J{\"a}rvel{\"a}}, {Berton}, \&
  {Crepaldi}}]{2021jarvela2}
{J{\"a}rvel{\"a}}, E., {Berton}, M., \& {Crepaldi}, L. 2021, Frontiers in
  Astronomy and Space Sciences, 8, 147

\bibitem[{{J{\"a}rvel{\"a}} {et~al.}(2022){J{\"a}rvel{\"a}}, {Dahale},
  {Crepaldi}, {Berton}, {Congiu}, \& {Antonucci}}]{Jarvela22}
{J{\"a}rvel{\"a}}, E., {Dahale}, R., {Crepaldi}, L., {et~al.} 2022, \aap, 658,
  A12

\bibitem[{{J{\"a}rvel{\"a}} {et~al.}(2018){J{\"a}rvel{\"a}},
  {L{\"a}hteenm{\"a}ki}, \& {Berton}}]{Jarvela18b}
{J{\"a}rvel{\"a}}, E., {L{\"a}hteenm{\"a}ki}, A., \& {Berton}, M. 2018, \aap,
  619, A69

\bibitem[{{J{\"a}rvel{\"a}} {et~al.}(2017){J{\"a}rvel{\"a}},
  {L{\"a}hteenm{\"a}ki}, {Lietzen}, {Poudel}, {Hein{\"a}m{\"a}ki}, \&
  {Einasto}}]{2017jarvela1}
{J{\"a}rvel{\"a}}, E., {L{\"a}hteenm{\"a}ki}, A., {Lietzen}, H., {et~al.} 2017,
  \aap, 606, A9

\bibitem[{{Keel} {et~al.}(2006){Keel}, {White}, {Owen}, \& {Ledlow}}]{Keel06}
{Keel}, W.~C., {White}, Raymond~E., I., {Owen}, F.~N., \& {Ledlow}, M.~J. 2006,
  \aj, 132, 2233

\bibitem[{{Kellermann} {et~al.}(1989){Kellermann}, {Sramek}, {Schmidt},
  {Shaffer}, \& {Green}}]{1989kellermann1}
{Kellermann}, K.~I., {Sramek}, R., {Schmidt}, M., {Shaffer}, D.~B., \& {Green},
  R. 1989, \aj, 98, 1195

\bibitem[{{Kharb} {et~al.}(2008){Kharb}, {O'Dea}, {Baum}, {Daly}, {Mory},
  {Donahue}, \& {Guerra}}]{Kharb08}
{Kharb}, P., {O'Dea}, C.~P., {Baum}, S.~A., {et~al.} 2008, apjs, 174, 74

\bibitem[{{Klindt} {et~al.}(2019){Klindt}, {Alexander}, {Rosario}, {Lusso}, \&
  {Fotopoulou}}]{Klindt19}
{Klindt}, L., {Alexander}, D.~M., {Rosario}, D.~J., {Lusso}, E., \&
  {Fotopoulou}, S. 2019, \mnras, 488, 3109

\bibitem[{{Komossa} {et~al.}(2006){Komossa}, {Voges}, {Xu}, {Mathur}, {Adorf},
  {Lemson}, {Duschl}, \& {Grupe}}]{2006komossa1}
{Komossa}, S., {Voges}, W., {Xu}, D., {et~al.} 2006, \aj, 132, 531

\bibitem[{{Kormendy} \& {Gebhardt}(2001)}]{2001kormendy1}
{Kormendy}, J. \& {Gebhardt}, K. 2001, in American Institute of Physics
  Conference Series, Vol. 586, 20th Texas Symposium on relativistic
  astrophysics, ed. J.~C. {Wheeler} \& H.~{Martel}, 363--381

\bibitem[{{Kotilainen} {et~al.}(2005){Kotilainen}, {Hyv{\"o}nen}, \&
  {Falomo}}]{Kotilainen05}
{Kotilainen}, J.~K., {Hyv{\"o}nen}, T., \& {Falomo}, R. 2005, aap, 440, 831

\bibitem[{{Kotilainen} {et~al.}(2016){Kotilainen}, {Le{\'o}n-Tavares},
  {Olgu{\'{\i}}n-Iglesias}, {Baes}, {An{\'o}rve}, {Chavushyan}, \&
  {Carrasco}}]{2016kotilainen1}
{Kotilainen}, J.~K., {Le{\'o}n-Tavares}, J., {Olgu{\'{\i}}n-Iglesias}, A.,
  {et~al.} 2016, \apj, 832, 157

\bibitem[{{Krongold} {et~al.}(2001){Krongold}, {Dultzin-Hacyan}, \&
  {Marziani}}]{2001krongold1}
{Krongold}, Y., {Dultzin-Hacyan}, D., \& {Marziani}, P. 2001, \aj, 121, 702

\bibitem[{{L{\"a}hteenm{\"a}ki} {et~al.}(2018){L{\"a}hteenm{\"a}ki},
  {J{\"a}rvel{\"a}}, {Ramakrishnan}, {Tornikoski}, {Tammi}, {Vera}, \&
  {Chamani}}]{2018lahteenmaki1}
{L{\"a}hteenm{\"a}ki}, A., {J{\"a}rvel{\"a}}, E., {Ramakrishnan}, V., {et~al.}
  2018, \aap, 614, L1

\bibitem[{{Magorrian} {et~al.}(1998){Magorrian}, {Tremaine}, {Richstone},
  {Bender}, {Bower}, {Dressler}, {Faber}, {Gebhardt}, {Green}, {Grillmair},
  {Kormendy}, \& {Lauer}}]{Magorrian98}
{Magorrian}, J., {Tremaine}, S., {Richstone}, D., {et~al.} 1998, aj, 115, 2285

\bibitem[{{Mao} {et~al.}(2018){Mao}, {Blanchard}, {Owen}, {Sjouwerman},
  {Singh}, {Scaife}, {Paragi}, {Norris}, {Momjian}, {Johnson}, \&
  {Browne}}]{Mao18}
{Mao}, M.~Y., {Blanchard}, J.~M., {Owen}, F., {et~al.} 2018, \mnras, 478, L99

\bibitem[{{Mathur} {et~al.}(2012){Mathur}, {Fields}, {Peterson}, \&
  {Grupe}}]{2012mathur1}
{Mathur}, S., {Fields}, D., {Peterson}, B.~M., \& {Grupe}, D. 2012, \apj, 754,
  146

\bibitem[{{Mathur} {et~al.}(2001){Mathur}, {Kuraszkiewicz}, \&
  {Czerny}}]{2001mathur1}
{Mathur}, S., {Kuraszkiewicz}, J., \& {Czerny}, B. 2001, NewA, 6, 321

\bibitem[{{Morganti}(2017)}]{Morganti17}
{Morganti}, R. 2017, Frontiers in Astronomy and Space Sciences, 4, 42

\bibitem[{{Olgu{\'\i}n-Iglesias} {et~al.}(2020){Olgu{\'\i}n-Iglesias},
  {Kotilainen}, \& {Chavushyan}}]{2020olguiniglesias1}
{Olgu{\'\i}n-Iglesias}, A., {Kotilainen}, J., \& {Chavushyan}, V. 2020, \mnras,
  492, 1450

\bibitem[{{Olgu{\'{\i}}n-Iglesias} {et~al.}(2017){Olgu{\'{\i}}n-Iglesias},
  {Kotilainen}, {Le{\'o}n Tavares}, {Chavushyan}, \&
  {A{\~n}orve}}]{2017olguiniglesias1}
{Olgu{\'{\i}}n-Iglesias}, A., {Kotilainen}, J.~K., {Le{\'o}n Tavares}, J.,
  {Chavushyan}, V., \& {A{\~n}orve}, C. 2017, \mnras, 467, 3712

\bibitem[{{Orban de Xivry} {et~al.}(2011){Orban de Xivry}, {Davies},
  {Schartmann}, {Komossa}, {Marconi}, {Hicks}, {Engel}, \&
  {Tacconi}}]{2011orbandexivry1}
{Orban de Xivry}, G., {Davies}, R., {Schartmann}, M., {et~al.} 2011, \mnras,
  417, 2721

\bibitem[{{Osterbrock} \& {Pogge}(1985)}]{1985osterbrock1}
{Osterbrock}, D.~E. \& {Pogge}, R.~W. 1985, \apj, 297, 166

\bibitem[{{Padovani}(2017)}]{2017padovani1}
{Padovani}, P. 2017, Nature Astronomy, 1, 0194

\bibitem[{{Paliya} {et~al.}(2018){Paliya}, {Ajello}, {Rakshit}, {Mandal},
  {Stalin}, {Kaur}, \& {Hartmann}}]{2018paliya1}
{Paliya}, V.~S., {Ajello}, M., {Rakshit}, S., {et~al.} 2018, \apjl, 853, L2

\bibitem[{{Peng} {et~al.}(2010){Peng}, {Ho}, {Impey}, \& {Rix}}]{Peng10}
{Peng}, C.~Y., {Ho}, L.~C., {Impey}, C.~D., \& {Rix}, H.-W. 2010, \aj, 139,
  2097

\bibitem[{{Peterson}(1997)}]{Peterson97}
{Peterson}, B.~M. 1997, {An Introduction to Active Galactic Nuclei}

\bibitem[{{Peterson}(2011)}]{2011peterson1}
{Peterson}, B.~M. 2011, ArXiv e-prints

\bibitem[{{Planck Collaboration} {et~al.}(2016){Planck Collaboration}, {Ade},
  {Aghanim}, {Arnaud}, {Ashdown}, {Aumont}, {Baccigalupi}, {Banday},
  {Barreiro}, {Bartlett}, {Bartolo}, {Battaner}, {Battye}, {Benabed},
  {Beno{\^\i}t}, {Benoit-L{\'e}vy}, {Bernard}, {Bersanelli}, {Bielewicz},
  {Bock}, {Bonaldi}, {Bonavera}, {Bond}, {Borrill}, {Bouchet}, {Boulanger},
  {Bucher}, {Burigana}, {Butler}, {Calabrese}, {Cardoso}, {Catalano},
  {Challinor}, {Chamballu}, {Chary}, {Chiang}, {Chluba}, {Christensen},
  {Church}, {Clements}, {Colombi}, {Colombo}, {Combet}, {Coulais}, {Crill},
  {Curto}, {Cuttaia}, {Danese}, {Davies}, {Davis}, {de Bernardis}, {de Rosa},
  {de Zotti}, {Delabrouille}, {D{\'e}sert}, {Di Valentino}, {Dickinson},
  {Diego}, {Dolag}, {Dole}, {Donzelli}, {Dor{\'e}}, {Douspis}, {Ducout},
  {Dunkley}, {Dupac}, {Efstathiou}, {Elsner}, {En{\ss}lin}, {Eriksen},
  {Farhang}, {Fergusson}, {Finelli}, {Forni}, {Frailis}, {Fraisse},
  {Franceschi}, {Frejsel}, {Galeotta}, {Galli}, {Ganga}, {Gauthier}, {Gerbino},
  {Ghosh}, {Giard}, {Giraud-H{\'e}raud}, {Giusarma}, {Gjerl{\o}w},
  {Gonz{\'a}lez-Nuevo}, {G{\'o}rski}, {Gratton}, {Gregorio}, {Gruppuso},
  {Gudmundsson}, {Hamann}, {Hansen}, {Hanson}, {Harrison}, {Helou},
  {Henrot-Versill{\'e}}, {Hern{\'a}ndez-Monteagudo}, {Herranz}, {Hildebrandt},
  {Hivon}, {Hobson}, {Holmes}, {Hornstrup}, {Hovest}, {Huang}, {Huffenberger},
  {Hurier}, {Jaffe}, {Jaffe}, {Jones}, {Juvela}, {Keih{\"a}nen}, {Keskitalo},
  {Kisner}, {Kneissl}, {Knoche}, {Knox}, {Kunz}, {Kurki-Suonio}, {Lagache},
  {L{\"a}hteenm{\"a}ki}, {Lamarre}, {Lasenby}, {Lattanzi}, {Lawrence}, {Leahy},
  {Leonardi}, {Lesgourgues}, {Levrier}, {Lewis}, {Liguori}, {Lilje},
  {Linden-V{\o}rnle}, {L{\'o}pez-Caniego}, {Lubin}, {Mac{\'\i}as-P{\'e}rez},
  {Maggio}, {Maino}, {Mandolesi}, {Mangilli}, {Marchini}, {Maris}, {Martin},
  {Martinelli}, {Mart{\'\i}nez-Gonz{\'a}lez}, {Masi}, {Matarrese}, {McGehee},
  {Meinhold}, {Melchiorri}, {Melin}, {Mendes}, {Mennella}, {Migliaccio},
  {Millea}, {Mitra}, {Miville-Desch{\^e}nes}, {Moneti}, {Montier}, {Morgante},
  {Mortlock}, {Moss}, {Munshi}, {Murphy}, {Naselsky}, {Nati}, {Natoli},
  {Netterfield}, {N{\o}rgaard-Nielsen}, {Noviello}, {Novikov}, {Novikov},
  {Oxborrow}, {Paci}, {Pagano}, {Pajot}, {Paladini}, {Paoletti}, {Partridge},
  {Pasian}, {Patanchon}, {Pearson}, {Perdereau}, {Perotto}, {Perrotta},
  {Pettorino}, {Piacentini}, {Piat}, {Pierpaoli}, {Pietrobon}, {Plaszczynski},
  {Pointecouteau}, {Polenta}, {Popa}, {Pratt}, {Pr{\'e}zeau}, {Prunet},
  {Puget}, {Rachen}, {Reach}, {Rebolo}, {Reinecke}, {Remazeilles}, {Renault},
  {Renzi}, {Ristorcelli}, {Rocha}, {Rosset}, {Rossetti}, {Roudier},
  {Rouill{\'e} d'Orfeuil}, {Rowan-Robinson}, {Rubi{\~n}o-Mart{\'\i}n},
  {Rusholme}, {Said}, {Salvatelli}, {Salvati}, {Sandri}, {Santos},
  {Savelainen}, {Savini}, {Scott}, {Seiffert}, {Serra}, {Shellard}, {Spencer},
  {Spinelli}, {Stolyarov}, {Stompor}, {Sudiwala}, {Sunyaev}, {Sutton},
  {Suur-Uski}, {Sygnet}, {Tauber}, {Terenzi}, {Toffolatti}, {Tomasi},
  {Tristram}, {Trombetti}, {Tucci}, {Tuovinen}, {T{\"u}rler}, {Umana},
  {Valenziano}, {Valiviita}, {Van Tent}, {Vielva}, {Villa}, {Wade}, {Wandelt},
  {Wehus}, {White}, {White}, {Wilkinson}, {Yvon}, {Zacchei}, \&
  {Zonca}}]{Planck16}
{Planck Collaboration}, {Ade}, P.~A.~R., {Aghanim}, N., {et~al.} 2016, \aap,
  594, A13

\bibitem[{Plewa {et~al.}(1997)Plewa, Marti, Mueller, Rozyczka, \&
  Sikora}]{plewa97}
Plewa, T., Marti, J.~M., Mueller, E., Rozyczka, M., \& Sikora, M. 1997, Bending
  relativistic jets in AGNs

\bibitem[{{Pogge}(2011)}]{2011pogge1}
{Pogge}, R.~W. 2011, in Narrow-Line Seyfert 1 Galaxies and their Place in the
  Universe, \url{https://pos.sissa.it/126/002/pdf}

\bibitem[{{Rakshit} {et~al.}(2018){Rakshit}, {Stalin}, {Hota}, \&
  {Konar}}]{Rakshit18}
{Rakshit}, S., {Stalin}, C.~S., {Hota}, A., \& {Konar}, C. 2018, apj, 869, 173

\bibitem[{{Rau} \& {Cornwell}(2011)}]{Rau_Cornwell11}
{Rau}, U. \& {Cornwell}, T.~J. 2011, aap, 532, A71

\bibitem[{{Richards} \& {Lister}(2015)}]{2015richards1}
{Richards}, J.~L. \& {Lister}, M.~L. 2015, \apjl, 800, L8

\bibitem[{{Salucci} {et~al.}(2000){Salucci}, {Szuszkiewicz}, {Monaco}, \&
  {Danese}}]{Salucci00}
{Salucci}, P., {Szuszkiewicz}, E., {Monaco}, P., \& {Danese}, L. 2000, mnras,
  311, 448

\bibitem[{{Savolainen} {et~al.}(2006){Savolainen}, {Wiik}, {Valtaoja},
  {Kadler}, {Ros}, {Tornikoski}, {Aller}, \& {Aller}}]{Savolainen06}
{Savolainen}, T., {Wiik}, K., {Valtaoja}, E., {et~al.} 2006, \apj, 647, 172

\bibitem[{{Urry} \& {Padovani}(1995)}]{1995urry1}
{Urry}, C.~M. \& {Padovani}, P. 1995, \pasp, 107, 803

\bibitem[{{Urry}(2003)}]{2003urry1}
{Urry}, M. 2003, in Astronomical Society of the Pacific Conference Series, Vol.
  290, Active Galactic Nuclei: From Central Engine to Host Galaxy, ed.
  S.~{Collin}, F.~{Combes}, \& I.~{Shlosman}, 3

\bibitem[{{Vietri} {et~al.}(2018){Vietri}, {Berton}, {Ciroi}, {Congiu}, {Chen},
  {Cracco}, {Cattapan}, {Frezzato}, {La Mura}, {Peruzzi}, \&
  {Rafanelli}}]{vietri18}
{Vietri}, A., {Berton}, M., {Ciroi}, S., {et~al.} 2018, in Revisiting
  Narrow-Line Seyfert 1 Galaxies and their Place in the Universe, 47

\bibitem[{{Wiegert} {et~al.}(2015){Wiegert}, {Irwin}, {Miskolczi}, {Schmidt},
  {Mora}, {Damas-Segovia}, {Stein}, {English}, {Rand}, {Santistevan},
  {Walterbos}, {Krause}, {Beck}, {Dettmar}, {Kepley}, {Wezgowiec}, {Wang},
  {Heald}, {Li}, {MacGregor}, {Johnson}, {Strong}, {DeSouza}, \&
  {Porter}}]{Wiegert15}
{Wiegert}, T., {Irwin}, J., {Miskolczi}, A., {et~al.} 2015, \aj, 150, 81

\end{thebibliography}

\appendix

\section{NIR/optical analysis}
\subsection{NIR}
\label{app:imaging-main}
To perform the photometric decomposition of the $Ks$-band image, we used GALFIT version 3 \citep{Peng10}, which allows simultaneous fitting of several components that contribute to the total 2D light distribution of the source. To properly model the AGN, a good PSF model is needed. GALFIT is capable of extracting the PSF directly from the image of a star once it is centred and sky-subtracted. This allows more accurate PSF modelling since no analytical functions are needed.

We estimated the zeropoint of the image and its associated error by using stars in the field of view with 2MASS magnitudes in $Ks$-band and applying the aperture photometry technique. The image was sky-subtracted and the sky error was estimated in several separate regions of 100$\times$100~px. We estimated the sky error effect by fitting with the $\pm$1$\sigma$ values. For magnitudes an additional zeropoint estimate error was added. 

We modelled the AGN component with a PSF. We tried with several approaches using IRAF and also with different PSF models created in GALFIT. First of all, we used the PSF photometry technique in IRAF. We did several attempts using the \textit{daophot} package: we started creating a stellar PSF with eleven stars with known 2MASS magnitudes, but more stars were needed to obtain a good PSF. Then we did the same procedure as before using multiple stars but the PSF obtained was still affected by the distortions in the detector along the field of view. To fix this problem, we modelled the PSF shape as a linear function of the position on the image and we also chose a bigger value for the radius of PSF model (fifteen times the mean FWHM) to see the shape of the stars' wings. This model was still affected by errors due to the field being too crowded. For these reasons, we tried taking the average FWHM of the stars in the image and creating a Gaussian model using GALFIT, and then used it as the PSF model. The PSF size was indeed too small to let the wings be visible. Then, we tried using a nearby star as close as possible to the AGN to minimise the effects of the position-dependent PSF. We extracted the isolated star pretty close to the galaxy and we used this star as the model PSF itself. This attempt was not optimal since the star was quite faint compared to the AGN. We choose to model the nearby star to smooth out any tiny imaging errors affecting the star. We fitted the nearby star with one Gaussian in GALFIT but only one function was not enough to properly model the PSF. For this reason, we tried fitting the star in GALFIT with different combinations of Gaussians and exponential disks and extracting the model PSF. The resulting PSF used for the galaxy fit did not give a proper model for the host galaxy. Finally, we obtained the best PSF model fitting five Sérsic functions for the nearby star, freezing the sky mean value to zero in the fit. 
 
The galaxy modelling was started by fitting only the PSF and then adding more components one at a time as required. The residuals were checked after every fit, and the component parameters were varied to ensure the stability of the result. We fixed the central coordinates of the AGN component but kept other parameters free. After achieving a good fit, determined by the reduced $\chi^2_{\nu}$, and the visual inspection of the model and the residuals, we also visually checked all the subcomponents to confirm they looked physically reasonable. 

The S\'{e}rsic profile was used to model the various components of our sources:

\begin{equation}
    I(r) = I_e \texttt{exp} \Bigg[ -\kappa_n \Bigg( \bigg( \frac{r}{r_e} \bigg)^{1/n} -1 \Bigg) \Bigg]
,\end{equation}

where $I(r)$ is the surface brightness at radius $r$, $\kappa_n$ is a parameter connected to the S\'{e}rsic index, $n$, so that $I_e$ is the surface brightness at the half-light radius, $r_e$ \citep{2005graham1}. By changing the S\'{e}rsic index, $n$, the S\'{e}rsic profile can be used to model varying light distributions in galaxies, for example, classical and pseudo-bulges, and early- and late-type morphologies. Smaller values of $n$ ($\lesssim$ 2) are associated with galaxies with late-type morphology and pseudo-bulges, and larger values of $n$ ($\gtrsim$ 4) with elliptical galaxies and classical bulges \citep{2005graham1}. 

After a good fit was achieved we extracted the radial surface brightness profile from the observed image, the model image, and the separate component images using IRAF task $ELLIPSE$, which fits concentric elliptical isophotes to a 2D image. For the original image and the model image we used similar $ELLIPSE$ parameters to get comparable fits. For the individual components, we took the values for the central coordinates, the axial ratio, and the position angle from the GALFIT best-fit parameters. The error estimate takes into account the most important error sources, the sky value error, and the zeropoint error.

\end{document}